\documentclass[12 pt]{article}
\usepackage{appendix}
\usepackage{authblk}

\begin{document}

\font\tenscr   = rsfs10 at 12pt 
\font\sevenscr = rsfs7  at 9pt  
\font\fivescr  = rsfs5  at 7pt  

\skewchar\tenscr   = '177
\skewchar\sevenscr = '177
\skewchar\fivescr  = '177

\newfam\scrfam

\textfont\scrfam         = \tenscr
\scriptfont\scrfam       = \sevenscr
\scriptscriptfont\scrfam = \fivescr

\def\scr{\fam\scrfam}

\title{Siegert State Approach to Quantum Defect Theory}

\author[1]{C. Hategan}
\author[1]{R.A. Ionescu}
\author[2]{H.H. Wolter}
\affil[1]{\small Institute of Atomic Physics, Bucharest, P.O.Box  MG-6, Romania}
\affil[2]{\small Fakult\"at f\"ur Physik, Universit\"at M\"unchen, Am Coulombwall 1, D-85748 Garching, Germany}

\date{}

\maketitle

\begin{abstract}
The Siegert states are approached in framework of Bloch-Lane-Robson formalism for
 quantum collisions. The Siegert state is not described by a pole of Wigner R- matrix
but rather by the equation $1- R_{nn}L_n = 0$, relating R- matrix element $R_{nn}$
to decay channel logarithmic derivative $L_n$. Extension of Siegert state equation to
multichannel system results into replacement of channel R- matrix element $R_{nn}$ by its
reduced counterpart ${\cal R}_{nn}$. One proves the Siegert state is a pole, 
$(1 - {\cal R}_{nn} L_{n})^{-1}$, of
multichannel collision matrix. The Siegert equation $1 - {\cal R}_{nn} L_{n} = 0$, ($n$ - Rydberg
channel), implies basic results of Quantum Defect Theory as Seaton's theorem,
complex quantum defect, channel resonances and threshold continuity of averaged
multichannel collision matrix elements.
\end{abstract}

\newpage
\section{Introduction}

Quasistationary states with complex eigenenergies have been introduced as an important
concept in theories of quantum collisions and decay. They were firstly
introduced by Gamow as complex-energy eigenstates in order to describe radioactive
decay, associating the lifetime of a quasistationary state with the imaginary part of
its energy. Thereafter Kapur and Peierls introduced a discrete set of complex energy
eigenstates depending on the scattering energy. Siegert \cite{siegert1939}
introduced a class
of solutions of the Schroedinger equation satisfying regularity conditions at the origin
and an outgoing-wave boundary conditions at infinity. The Siegert solution is given as
a discrete set of complex momenta $k_{\lambda}$ which are the poles of the collision matrix
in the complex $k$- plane. Pure positive imaginary momenta represent bound states,
pure negative imaginary ones virtual states, and poles lying close but below the real
positive $k$- axis quasistationary or resonant states. The Siegert states provide a unified
description of bound and resonant states.

Gamow-Siegert states were and are an important ingredient in quasistationary
formalisms, {\it e.g.} \cite{perelomov1998},
in Green function formalisms, {\it e.g.} \cite{kukulin1989},
or in the Rigged Hilbert Space formulation,
{\it e.g.} \cite{hernandez2003,michel2009}. 

The present approach to Siegert state is developed in framework of Bloch-Lane-Robson
formalism for quantum collisions, \cite{lane1966,robson1967},
using Bloch boundary condition operator $\cal L$, \cite{bloch1957}. 
The boundary condition should
be congruent with outgoing behaviour at infinity of the Siegert state. The usual
Schroedinger equation and Bloch $out$ wave operator result into an eigenvalue equation
for Siegert state. The Schroedinger equation for Siegert state is rewritten in channel
space in terms of scattering operators; it will result into an equation, $1- {R}_{nn}L_n=0$,
relating R- matrix element to logarithmic derivative of decay channel. Implication of
(bound or quasistationary) Siegert state equation in multichannel collisions results into
a pole, $(1- {\cal R}_{nn}L_n)^{-1}$, of collision matrix. The Siegert state equation, in its R-matrix
parametrization, is applied to electron scattering on atoms resulting in a new approach
to Multichannel Quantum Defect Theory. 

\section{Siegert State in Collision Theory}

In this section we follow the formalism of Lane and Thomas \cite{lane1958} and
Lane and Robson \cite{lane1966,robson1967}.
The Bloch outgoing-wave boundary condition operator, for
short the Bloch operator, is defined by
\begin{eqnarray}
{\cal L} = \sum_c |c> \frac {\hbar^2} {2 m_c} \delta (r_c - a_c)[ \frac {d} {d r_c} - \frac {b_c -1} {a_c}] <c|
\end{eqnarray}
\begin{eqnarray}
b_c =  { a_c \frac {d O_c} {dr}}/ O_c
\end{eqnarray}
where $|c>$  and $O_c$ denote a channel state and the channel outgoing wave,  $r_c$ and $m_c$
the channel coordinate and reduced mass, and $a_c$ the channel radius, respectively. This
Bloch operator ${\cal L}$ projects out outgoing-waves states in all channels provided the set
$|c>$ is orthogonal.

Given a hamiltonian $H$, we define an ideal state $|\lambda >$ by, \cite{robson1967},
\begin{eqnarray}
 (H-E_{\lambda}) |\lambda > = 0,
\end{eqnarray}
{\it eg} an R-matrix state, subject to mixing with complementary states due to changes in
boundary conditions and interactions. The Siegert state  $|g_{\lambda}>$, on the other hand, as
a purely outgoing one, is an eigenstate of the Bloch operator with zero eigenvalue,
${\cal L} |g_{\lambda}> = 0$, \cite{robson1975}. It obeys a modified Schroedinger equation, \cite{robson1967},
\begin{eqnarray}
(H + {\cal L} - E)|g_{\lambda}>=(E_{\lambda} + {\Phi}_{\lambda} - E) | \lambda>,
\end{eqnarray}
with ${\Phi}_{\lambda}$ a quantity defining a complex level shift. 
Inserting it into the usual Schroedinger equation
\begin{eqnarray}
(H + {\cal L}(E^{'}) - E^{'}) |g_{\lambda}> = {\cal L}(E^{'}) |g_{\lambda}> = 0
\end{eqnarray}
one obtains a constraint for the energy of the quasistationary state $|g_{\lambda}>$, namely
$ E^{'} = {\cal E}_{\lambda} = E_{\lambda} + \Phi_{\lambda}$.

We also define the corresponding Green operator for the Hamiltonian $H$ and
boundary condition ${\cal L}$ as, \cite{robson1969}
\begin{eqnarray}
{\cal G} = (H + {\cal L} -E)^{-1}.
\end{eqnarray}
The Siegert state equation 
{${\cal G}^{-1}|g_{\lambda}>=0 $}
is rewritten in channel space by projecting on the adjunct
state $<\tilde{g}_{\lambda}|$ and by inserting the channel projector $\sum_c |c><c|$, assuming that the
channels $c$ exhaust the decay of Siegert state,
\begin{eqnarray}
\sum_{c c^{'}} \tilde {\omega}_{\lambda c}
<c|{\cal G}^{-1}({\cal E}_{\lambda})|c^{'}> \omega_{c^{'} \lambda}= 0 \nonumber
\end{eqnarray}
with $\omega$ as the Siegert state reduced widths.

The Green operator ${\cal G}$ defines the ${\scr R}$ matrix, \cite{robson1969},
\begin{eqnarray}
{\scr R}_{c c^{'}} = < c^{'}|{\cal G}| c>
\end{eqnarray}
associated with collision matrix $U$,
\begin{eqnarray}
U = \Omega W \Omega = \Omega (1 + 2i P^{\frac {1} {2}} 
{\scr R} 
P^{\frac {1} {2}}) \Omega
\end{eqnarray}
where $\Omega$ and $P$ are the phase and penetration factor diagonal matrices \cite{lane1958}. The Siegert
state equation in terms of the ${\scr R}$ 
matrix
becomes
\begin{eqnarray}
\sum_{c c^{'}} \tilde {\omega}_{\lambda c} {\scr R}^{-1}_{c c^{'}} \omega_{c^{'} \lambda} =0 
\end{eqnarray}

The Green operator is related to the R- matrix, as, \cite{lane1966},
\begin{eqnarray}
{\scr R}_{c c^{'}}= [(1- RL)^{-1}R]_{c c^{'}},
\end{eqnarray}
with $L$ as the channel logarithmic derivative. The Siegert state equation in R- matrix
terms becomes
\begin{eqnarray}
\sum_{c^{'}} (1 - RL)_{c c^{'}} \omega_{c^{'} \lambda} = 0 
\end{eqnarray}
The poles ${\cal E}_{\lambda}$ are the complex roots of the determinant equation
\begin{eqnarray}
|1 - R({\cal E}_{\lambda}) L({\cal E}_{\lambda})|  = 0 
\end{eqnarray}
Both $R$ and $L$ are energy dependent and are assumed to be analytically continuable to
complex ${\cal E}_{\lambda}$.

We now consider the case of Siegert state $g_{\lambda}$ in a particular channel $c=n$. Coupling
to other channels is taken into account in terms of the multi-channel reduced R- matrix
element ${\cal R}$,
\begin{eqnarray}
{\scr R}_{n n} = [(1- RL)^{-1}R]_{n n} = (1-{\cal R}_{n n} L_n)^{-1}{\cal R}_{n n}.
\end{eqnarray}
{It is a complex quantity due to implication of open channels logarithmic derivatives 
in its very definition \cite{lane1958}.}
Accordingly, the single channel equation for the quasistationary level $g_{\lambda}$ in channel $n$
becomes
\begin{eqnarray}
[1 - {\cal R}_{n n}({\cal E}_{\lambda}) L_n({\cal E}_{\lambda})]  \omega_{n \lambda } = 0 
\end{eqnarray}
The complex energy pole ${\cal E}_{\lambda n}$ is again given by the implicit equation 
$1 - {\cal R}_{n n}({\cal E}_{\lambda}) L_n({\cal E}_{\lambda})  = 0$. 

To summarize this section, the matching of the R- matrix element $R_{n n}$ to the channel
logarithmic derivative $L_n$ is expressed by the channel equation $1-R_{n n} L_n = 0$. Below
threshold, with $L_n =S_n$
{($S_n$ - shift function)}, 
this is the R- matrix equation for bound states, $1-R_{n n} S_n = 0$
or $R_{n n}^{-1}= S_n$: a bound state appears at that energy at which the internal $R_{n n}^{-1}$ 
and the external $S_n$ logarithmic derivatives match. At positive energy it is the
logarithmic derivative $L_n$ of the outgoing wave which has to match the logarithmic
derivative of internal wave function $R_{n n}^{-1}$ at the channel radius. This condition defines
the quasistationary state. The root of the implicit equation 
$1 - R_{n n}({\cal E}_{\lambda}) L_n({\cal E}_{\lambda})  = 0$ is
a complex energy pole ${\cal E}_{\lambda}$. As in the case of bound states, the equation yields a set of
eigenenergies which now are complex. Thus the channel equation $1-R_{n n} L_n = 0$ defines
both the bound state (below threshold) or quasistationary state (above threshold).

The R- matrix Siegert  equation in case of multichannel systems becomes 
$1 - {\cal R}_{n n}({\cal E}_{\lambda}) L_n({\cal E}_{\lambda})  = 0$
where ${\cal R}_{n n}$ is reduced R- matrix element. The essential term in the collision
matrix, describing effect of the eliminated channels on channel $n$, is $(1-{\cal R}_{n n}L_n)^{-1}$, both
below and above $n$-channel threshold. The collision matrix formalism, expresses here
in the R- matrix approach in terms of Siegert states, is applied in the next chapter to
electron scattering on atoms.

\section{Siegert State and   Quantum Defect Theory}

The  Siegert R- matrix equation, $L_n^{-1} - {\cal R}_{nn} = 0$, can be discussed in two different aspects, either
with respect to the R-matrix element or to the logarithmic derivative. 
The logarithmic
derivative can have a resonant form, {\it eg} as proposed by Abramovich {\it et al} 
\cite{abramovich1992},
resulting
in a generalization of the 
{Wigner-Breit-Baz threshold}
cusp theory. In atomic physics an energy-dependent
logarithmic derivative with poles on the real axis is used for studying Rydberg states,
at negative energy. By applying the Siegert equation to electron Rydberg states, we
derive in a new way basic results of Quantum Defect Theory.

This approach to QDT, based essentially on Siegert equation, is an alternative to
classical variant. It follows the spirit of previous approaches to MQDT based on R-Matrix
\cite{lane1986}
or on Level- matrix. The Siegert equation implies Seaton's theorem
both for one-channel and multichannel problems, relating the complex quantum defect to
complex scattering phase-shift. Thereafter the Level- matrix approach is reformulated
in order to relate (multichannel) collision matrix to Siegert equation. The MQDT
collision matrix, in its relation to Siegert equation, is applied to derivation of channel
resonances and to prove the Threshold Continuity Theorem. This theorem relates the
collision matrix elements above threshold to the averaged ones below threshold; it is
an alternative to Gailitis' theorem.

\subsection{Quantum Defect and Siegert Equation}

The logarithmic derivative of a purely Coulombic electron state ($n \equiv e$;
{$e$ Rydberg channel label}) for energy $E$
below the threshold energy $E_{\pi}$  is given by,  \cite{baz1971} p. 408, \cite{baz1971e}, \cite{landau1980} p. 708,
$L_e^< = - \cot \pi \sqrt {e^2_1 e^2_2  m / 2 \hbar^2 (E_{\pi} - E)}$  ($e_1$, $e_2$, and $m$ are the
particles electric charges and the reduced mass, respectively). The pure Coulomb states,
{\it i.e.} in the absence of an inner core, are defined by the level equation $(L_e^<)^{-1}=0$. It
yields the eigenenergies $E_{\pi} - E_n= (1/n^2)$ $e_1^2 e_2^2 m/ 2 \hbar^2$, with $n$ an integer number, 
the principal quantum number.

We now consider an inner electron core, characterized by a finite R-matrix element
$R_{ee}$, and a level equation $(L_e^<)^{-1} - R_{ee}= 0$. We are interested in Rydberg states,
which have a very large spatial extent relative to the core. Thus we are allowed to
use the same form of the logarithmic derivative as above. Then the effect of the inner
core on the electron spectrum (resulting into a level-shift) is expressed by replacing the
principal quantum number $n$ by an effective non-integral quantum number $\nu$, or a so-called
quantum defect $\mu$ defined by the relation $\nu = n - \mu$. The logarithmic derivative
below threshold then is written as $L_e^< = - \cot \pi \nu = \cot \pi \mu$ and the relation between the
eigenenergies and the effective quatum number is $E_{\pi}-E_{\nu}\propto 1/\nu^2$. The condition for
a single electron bound Rydberg state is now $\tan \pi \mu = R_{ee}$. It constitutes a relation
between the one-channel quantum defect $\mu$ to the Rydberg channel R-matrix element
$R_{ee}$.  Let proceed with Siegert equation to Seaton's theorem and to complex quantum defect.

The logarithmic derivative for an  attractive Coulomb channel above threshold is 
\cite{baz1971,baz1971e,landau1980}
$L_e^{>} = i$;  in zero-energy limit the channel penetration factor is
$P_e = 1$  and its shift-factor is $S_e = 0$. The equation for the scattering phase-shift $\delta_e$ in an
open electron channel given in R-matrix theory by $\tan \delta_e =
R_{ee} P_e /(1 - R_{ee} S_e)$, reduces
to $\tan \delta_e = R_{ee}$. Comparing to the above relation for a Rydberg state we obtain Seaton
theorem, $\delta_e = \pi \mu$, relating the scattering phase shift above threshold to the quantum
defect of the spectrum below threshold. For a single channel the quantum defect and
the phase shift are real.

In a multi-channel situation with a coupling of the closed channel $e$ to other closed
and open channels the reduced R- matrix element ${\cal R}_{ee}$ is complex. Thus the Siegert equation becomes
\begin{eqnarray}
{\cal R}_{ee} =\tan \pi {\tilde \mu} = \tan {\tilde \delta}_e, 
\end{eqnarray}
with complex quantum defect, $\tilde \mu$, and scattering phase shift, $\tilde \delta_e$.
 The imaginary part of complex quantum defect $\pi Im \tilde \mu = Im ({\arctan} {\cal R}_{ee})$ is positive, 
provided $Im {\cal R}_{ee}$ is positive too. Actually for a multichannel system
 $Im {\cal R}_{ee} > 0$; resulting from relation of reduced R- matrix element, ${\cal R}_{ee}$, 
to subunitary value of collision matrix one, $ |W_{ee}| < 1$. 
 The change $\Delta \mu = \tilde \mu - \mu$ of the quantum defect in the limit when 
$\tan \pi \mu \simeq \pi \mu$ becomes $\pi \Delta \mu = {\cal R}_{ee} - R_{ee}$; 
it proves $ \Delta \mu$ originates in effective term of reduced R- matrix element (i.e. coupling of Rydberg channel 
to complementary ones).

\subsection{Multi-Channel Quantum Defect Equations}

The one-channel Quantum Defect Theory assumes the inner multielectron core is inert.
The only 'active' state is the Rydberg one, defining the 'Rydberg channel'. The Rydberg
state is a highly excited bound state, located just below the ionization threshold and its
eigenenergy is given by the quantum defect. When its energy rises above the threshold
it is transformed into a scattering state characterized by a scattering phase shift. (As
long as it is closely above threshold we will refer to it as a Rydberg state at positive
energy.) The quantum defect below threshold is related to the scattering phase shift
above threshold by Seaton's theorem. In the following we will denote the situation
when the Rydberd state is below threshold by the superscript $<$, and the one above
threshold by $>$.

In the Multi-Channel Quantum Defect Theory (MQDT) we consider that the multielectron
core can be excited, which generates several excited states, lower in energy than
the Rydberg state. As the Rydberg state is very close to threshold, we assume that these
excited states all form open channels. These then define the scattering channels, labelled
by $N$, to which the Rydberd state is coupled. Above the Rydberg channel threshold the
open channels thus consist both of the $N$ core-excited channels and the Rydberg channel
$n$. The $N + 1$ channel reaction system is described by the collision matrix $W^>$ with
components $W^>_N$, $W^>_{nn}$ and coupling terms $W^>_{Nn}$ and $W^>_{nN}$.  Below threshold only the
$N$ core-excited channels are open and the reaction system is described by the collision
matrix $W^<_N$, which, of course, depends also on the closed Rydberg channel. But the
essential difference between the situations of a closed and an open Rydberg channel
is the change in the logarithmic derivatives $L^<_n$ and $L^>_n$. Thus MQDT relates collision
matrix below threshold $W^<_N$ to the change across the threshold of the channel logarithmic
derivative, $\Delta L_n$ = $L_{n>}-L_{n<}$ and to the collision matrix elements above threshold 
$W^>_N$, $W^>_{nn}$, $W^>_{Nn}$, $W^>_{nN}$.

Assuming that the only changing parameter across the threshold is the $n$-channel
logarithmic derivative we obtain the following relation connecting the collision matrices
below and above threshold, provided $L_{n<}$ is real, \cite{hategan1995}
\begin{eqnarray}
 W_N^< = W^>_N -
W_{Nn}^>{{1}\over{-(\Delta L_{n})^{\star}/( \Delta L_{n}) +W^>_{nn}}}W_{nN}^>
\end{eqnarray}
The collision matrix $W_N^>$ above threshold is
\begin{eqnarray}
W_N^> &=& W^0_N +
W_{Nn}^>{{1}\over{-L_{n}^{> \star}/ L_{n}^> +W^>_{nn}}}W_{nN}^>
\end{eqnarray}
\begin{eqnarray}
W^>_{nn} &=& 1  + 2i P_n^{1/2} ({\cal R}_{nn}^{-1} - L_{n>})^{-1} P_n^{1/2} \nonumber \\
W^>_{nn} &=& L_{n}^{> \star}/ L_{n}^>  + 2i P_n (L_{n}^>)^{-2}(L_{n}^{> -1}-
{\cal R}_{nn})^{-1} \\
{\cal R}_{nn} &=& R_{nn} - R_{nN}(R_{NN} -L^{-1}_N)^{-1}R_{Nn} \nonumber \\
W^0_N &=&
1+2i P_N^{1/2}(R_N^{-1}-L_N)^{-1}P_N^{1/2} \nonumber
\end{eqnarray}
where ${\cal R}_{nn}$ is reduced R-matrix element of the  $n$ channel and
and $W^0_N$  is the collision
matrix for the $N$ uncoupled core-exited channels.

All $W^>$ Collision Matrix elements related to $n$-channel, 
$W^>_{nN}$, $W^>_{Nn}$ and $W^>_{nn}-L_{n}^{> \star}/ L_{n}^>$
are proportional to $({\cal R}_{nn}- L_{n}^{> -1})^{-1}$ and this results into same dependence 
for the increment $\Delta W^>_N=W^>_N-W^0_N$. 
\begin{eqnarray}
\Delta W^>_N= A_{Nn}(L_{n}^{> -1}-{\cal R}_{nn})^{-1}A_{nN}
\end{eqnarray}
The complementary matrix term $A_{Nn}=A_N R_{Nn}$  is not dependent on Siegert state
paramaters $L_n$ and ${\cal R}_{nn}$. The increment of collision matrix below threshold 
$\Delta W^<_N=W^<_N-W^0_N$
is proportional to $({\cal R}_{nn}- L_{n}^{< -1})^{-1}$ and has a similar dependence
\begin{eqnarray}
\Delta W^<_N= A_{Nn}(L_{n}^{< -1}-{\cal R}_{nn})^{-1}A_{nN}
\end{eqnarray}

The increments of the collision matrix below and above threshold $\Delta W^<_N$ and $\Delta W^>_N$ 
due to the couplings are related as
\begin{eqnarray}
\Delta W^<_N = \Delta W^>_N  {{L_{n>}^{-1}-{\cal R}_{nn}} \over {L_{n<}^{-1}-{\cal R}_{nn}} }
\end{eqnarray}
This relation involves Siegert equations below and above threshold. It gives rise to a
pole of collision matrix below threshold defined by the Siegert equation for the closed
channel 
$1/L_{n<} - {\cal R}_{nn}=0$. 

The above expressions give the MQDT equations for electron-atom collisions
provided the logarithmic derivative of the Rydberg channel $n \equiv e$ is explicitely 
specified in the threshold limit. Using the results for the pure Coulombic case we have
$L_e^< = - \cot \pi \nu$, 
$L_e^> =  i$,
$\Delta L_e = L_e^> - L_e^< =  e^{i \pi \nu}/\sin \pi \nu$,
$(\Delta L_e)^{*}/(\Delta L_e) =  e^{- 2i \pi \nu}$,
$\tau = Im \Delta L_e / Re \Delta L_e =  \tan \pi \nu$.
Note that the only quantity here, strongly dependent on
energy is the Rydberg channel logarithmic derivative $L_e^<$. We then obtain the MQDT
result for electron-atom scattering (see \cite{seaton1983,burke2011})
\begin{eqnarray}
W_N^< &=& W_N^>-
W_{Ne}^>{{1}\over{- e^{- 2i \pi \nu} +W^>_{ee}}}W_{eN}^>
\end{eqnarray}


 The poles of collision matrix are related either to R- matrix or to Siegert equation, (eq. 22). 
The R- matrix poles correspond to "inner resonances", originating in multielectron excitations of the inner core, 
\cite{lane1986}.
The poles related to Siegert equation, $(1/L_{e<} - {\cal R}_{ee})^{-1}$, describe the "channel resonances" 
in electron scattering on atoms and ions. The "channel resonances" originate in excitation of Rydberg far-away 
located states. The complex energies of "channel resonances"  are obtained either in terms of reduced R- matrix 
element  or complex quantum defect, 
(see \cite{seaton1983}),
or transition matrix element (see \cite{sobelman1981}).

For $s$-wave scattering on external (outside inner core) neutral fields, $L_{n>} = iP_n = i\rho$, 
$L_{n<} = S_{n<} = -\rho$,  $\Delta S_n=\rho$, ($\rho = k_n r$), ($k_n$ -wave number, $r$-channel radius), one
obtains $\tau=P_n /\Delta S_n = 1$ and $\arctan \tau = \pi /4$, (see \cite{fano1981}).
The increments in this case
are related by $\Delta W_N^< = \Delta W_N^>(i+ \rho {\cal R}_{nn})/(1 + \rho {\cal R}_{nn})$
 which in zero-energy limit of potential
scattering ($ \rho \to 0$) reduces to 
{Wigner-Breit-Baz threshold}
cusp theory result $\Delta W_N^< = i \Delta W_N^>$, 
\cite{baz1971}, ch.IX, \cite{baz1971e}.

\subsection{Threshold Continuity Theorem}

The collision matrix, both below and above $e$-threshold, is
\begin{eqnarray}
W_N = W^0_N + \Delta W_N
\end{eqnarray}
with $W^0_N$
as collision matrix for $N$ independent channels, (uncoupled to Rydberg one),
and $\Delta W_N$ as effective term due to channels couplings. According to (20-21) the
dependence on threshold channel of the effective term is
\begin{eqnarray}
\Delta W_N = A_{Ne} (L^{-1}_e - {\cal R}_{ee})^{-1}A_{eN}  
\end{eqnarray}
where for $\Delta W_N$ superscripts $>$ or $<$ one has to insert the corresponding logarithmic
derivatives $L^>_e$ or $L^<_e$ of the channel $e$, respectively. In following $\Delta W_N$ denotes only the
term $(L^{-1}_e - {\cal R}_{ee})^{-1}$; the complementary matrix term $A_{Ne}$ does not depend significantly
on energy.

The only term of collision matrix $\Delta W_N$ strong dependent on energy is the closed
Rydberg electron channel term $(L^{-1}_{e<} - {\cal R}_{ee})$. The energy dependence is contained in
logarithmic derivative; below threshold $L_{e<} = - \cot \pi \nu$; above threshold $L_{e>} = i$. The
${\cal R}_{ee}$ matrix element for multielectron inner core is considered as nearly constant in
threshold region. The energy average of the effective term below threshold 
$\overline {\Delta W_N^<}$
\begin{eqnarray}
\overline {\Delta W_N^<} = \int \rho(E) dE \Delta W_N^<
\end{eqnarray}
where averaging weight is $\rho(E)$ - density of states, 
\cite{baz1971}, p. 410, \cite{baz1971e},
\begin{eqnarray}
 \rho(E) = {1 \over 2} {e_1 e_2 \over \hbar} \sqrt {m \over 2} {1 \over (E_{\pi} - E)^{3/2}}
\end{eqnarray}
Introducing the variable $\lambda = L_e^<$ one obtains
\begin{eqnarray}
 {d \lambda \over 1 + \lambda^2} = \pi \rho(E) dE
\end{eqnarray}
\begin{eqnarray}
\overline {\Delta W_N^<} = \int {\rho(E)dE \Delta W_N^<} = {1 \over \pi} \int {d \lambda \over 1 +\lambda^2} 
\Delta W_N^<
\end{eqnarray}

The denominator $1 + \lambda^2$  has in upper half plane a pole at $\lambda = i$, implying
\begin{eqnarray}
\int \rho(E) dE = {1 \over \pi} \int { d \lambda \over 1 + \lambda^2} = 1
\end{eqnarray}
The term $\Delta W_N^<$ has a pole in lower half plane,
\begin{eqnarray}
\Delta W_N^< = (1/\lambda - {\cal R}_{ee})^{-1}
\end{eqnarray}
at $\lambda = 1/{\cal R}_{ee} = {\cal R}_{ee}^*/|{\cal R}_{ee}|^2$, {\it ie} $Im \lambda \sim - Im {\cal R}_{ee} < 0$, 
(because $Im {\cal R}_{ee} >0$). As $\Delta W_N^<$ has no pole in upper half plane, by using the residue's theorem
\begin{eqnarray}
\overline {\Delta W_N^<} = {1 \over \pi} \int {d \lambda \over 1 +\lambda^2} \Delta W_N^< =  \Delta W_N^< (\lambda = i)
\end{eqnarray}
\begin{eqnarray}
\overline{  (L_{e <}^{-1} - {\cal R}_{e e})^{-1}} = (1/i - {\cal R}_{ee})^{-1} = (1/L_e^> - {\cal R}_{ee})^{-1} 
\end{eqnarray}
one obtains that the averaged effective term and averaged reduced collision matrix, are
continous across threshold, 
$\overline{ \Delta W^<_N} = \Delta W^>_N$ and $\overline{W^<_N} = W^>_N$. The physical interpretation
of this result is: the near-threshold (high excited) bound states in coulombian field are
physically similar to scattering states (small positive energy) of continuum spectrum,
\cite{baz1971}, ch. IX, \cite{baz1971e}.
The energy average (over mean spacing of levels)
of the reaction total cross section of open channel $a \in N$, 
$\sigma^t_{aa} \sim Re (1 - W_{aa})$, results into Gailitis' theorem,
$\overline {\sigma^{<t}_{aa}} = \sigma^{>t}_{aa}$.

\subsection{R-matrix approach to MQDT; a remark}

The Multichannel Quantum Defect Theory (MQDT) is based on possibility of separating
the effects of long and short range interactions between an electron and an atomic core,
{\it eg} \cite{seaton1983,fano1986,burke2011}. 
The effects of short range interactions, within the core,
are very complex but, nevertheless, can be concisely represented by a global parameter,
named Quantum Defect. The long range interactions, (represented by simple fields as
the Coulomb or dipolar ones), are treated analitically by extensive use of Coulomb or
other special functions \cite{seaton1983,greene1982}. On the other hand the general assumptions of the MQDT
are similar to those of R- matrix theory, \cite{lane1986}.
Developing this idea and by
using only basic properties of Whittaker and Coulomb functions, Lane has extracted
MQDT from Wigner's R-matrix theory. A relationship between K- matrix, on one
side, and R-matrix, boundary condition parameters and Coulomb functions, on other
side, was established. This relation was then rewritten, by using specific boundary
conditions, in a K- matrix form of MQDT. The MQDT was also derived from the Level-
matrix parametrization of the collision matrix \cite{hategan1995}.
This last
approach proves that the essential aspects of the MQDT originate in variation across
threshold of the logarithmic derivative of the Rydberg channel. In this work the role of
Rydberg states (from closed channel) for producing resonant effects in the competing
open reaction channels of the multichannel system is pointed out, by relating the Level-
matrix approach to the Siegert state equation.

\section{Conclusions}
The Siegert state, defined as a channel single particle state subject of $out$ wave boundary
conditions, is described in terms of Bloch-Lane-Robson formalism for quantum collisions and,
thereafter, its equation is related to scattering operators. This way the Siegert state,
either bound or quasistationary, is approached in terms of R- matrix and channel
logarithmic derivative. It is not described by a Wigner R- matrix pole but rather by an
equation relating R- matrix to channel logarithmic derivative. The channel equation
\begin{eqnarray}
1-R_{nn}L_n=0
\end{eqnarray}
(with $L_n$ real for bound state and complex for quasistationary states), results into (real
or complex energy) poles of the collision matrix. If the channel under question is part
of a multichannel system then the channel R- matrix element $R_{nn}$ is replaced by its
reduced counterpart ${\cal R}_{nn}$.

The Siegert state is reflected in complementary channels of the reaction system as
a pole $(1 - {\cal R}_{nn}L_n)^{-1}$ in the effective term $\Delta W_N \sim (1 - {\cal R}_{nn}L_n)^{-1}$
of collision matrix.
The effective term of collision matrix, corresponding of reduced R- matrix, describes
the effect of unobserved (eliminated) channel on observed (retained) ones.

The collision matrix for case of electron closed channel is equivalent to QDT
equations. The electron channel equation 
$L^{-1}_{e<} - R_{ee}= 0$ or $\tan \pi \mu = R_{ee}$ is equivalent
to Seaton' theorem, relating at zero energy the quantum defect $\mu$ to scattering phase
shift  $\delta_e$. The Seaton' equation in multichannel system, 
$L^{-1}_{e<} - {\cal R}_{ee}= 0$ or   $tan \pi \tilde {\mu} = {\cal R}_{ee}$
results into complex quantum defect, $\tilde {\mu}$. The equation results also in channel resonances,
defining energies in term of complex quantum defect. The Siegert state, formally enviced
in effective term 
$\Delta W^<$ of collision matrix, is origin of the 'channel resonance', as distinct
from 'inner' R- matrix resonance. The energy averaged collision matrices, evaluated
below and above threshold, are related by a threshold continuity equation, 
$\overline { W_N^<} = W_N^>$,
which is alternative to Gailitis' theorem.

The present approach to Siegert state in multichannel scattering is applied
to physics of multichannel electron scattering; it does reproduce some results of
Multichannel Quantum Defect Theory, without pretence for substitution of MQDT
exhaustive derivation, but rather guided by it.

This work does follow the philosophy of Lane's paper \cite{lane1986}: a direct derivation
of MQDT starting with a basic theory or concept. An interesting problem which does not fit in streamline
of the work is comparison of the Siegert state approach to previous theoretical formalisms for multichannel electron scattering
(see however Appendix).


~
\newline
{\bf Appendix}
\begin{appendices}
\section{The Siegert state approach versus theoretical models for
 multichannel electron scattering: some remarks}


The Collision matrix $W^<_N$,   for case of eliminated closed ($<$)  $n-$channel,  is related to Collision matrix elements $W^>_N$, $W^>_{Nn}$, $W^>_{nn}$ of open ($>$) channel system and to jump across threshold of logarithmic derivative $\Delta L_n$. 
Actually this equation relates two reaction systems (denoted $<$ and $>$) which have same internal dynamics, (R- matrix), but differ in interaction in channel space, ($L^<_n$ and $L^>_n$). 

The effective terms $\Delta W_N^>$ and $\Delta W_N^<$ of the Collision matrix are a formal frame for description of threshold effects in quantum collisions. Actually they generalize the 
{Wigner-Breit-Baz threshold}
cusp theory \cite{baz1971,baz1971e},
by relating threshold effects to reaction dynamics.
The threshold cusp is obtained in limit of potential scattering, {\it ie} no Siegert pole near threshold.

The   Collision matrix $W^<_N$ and $\Delta L^*_{n}/\Delta L_{n}$
correspond, respectively,  to physical Scattering matrix and to 
  long range Scattering  matrix defined in \cite{aymar1996}. 
 By specializing to Coulomb interaction, $\Delta L^*_{n}/\Delta L_n = e^{- 2i \pi \nu}$, the Collision matrix equation of MQDT is derived \cite{seaton1983}.
 The threshold continuity theorem of energy-averaged Collision matrix for multichannel electron scattering is straightforward derived.

One can prove that the MQDT equation for Collision matrix $W^<_N$ 
is equivalent to  reduced R- matrix,
 $ {\cal R}^<_N = R_N - R_{N n}(R_{nn} - 1/L^<_n)^{-1} R_{n N}$. 
The R- matrix, subject of natural boundary conditions, results in a reduced K- matrix parametrization of  $W^<_N$  Collision- matrix with, 
$ {\cal K}^<_N = K_N - K_{Nn} (K_{nn} + \tau^<)^{-1} K_{nN} $
and  $\tau^< = ({Im} \Delta L_n )/({Re} \Delta L_n)$.
This formal result for short-range effective $K$- matrix is in  common with channel elimination methods in  MQD theories
\cite{seaton1983}
or reaction matrix projector method \cite{fano1986}
or phase-shifted MQDT \cite{aymar1996}.

The effective term $\Delta W_N$ of Collision matrix is related to Siegert equation.
The Siegert pole  describes the 'channel resonances' while R- matrix poles correspond to 'inner resonances'. In this work the two types of resonances are quite separated; however they could be formally mixed as in Lane's  work 
\cite{lane1986} or in MQD theories. 

The Siegert boundary condition is just condition that the Jost function expressing the fact that there is no incoming wave should vanish. 

The QD formalism, developed by  Greene, Rau  and Fano \cite{greene1982},
is based on Jost function.
The Jost matrices of Eigenchannel formalism \cite{fano1986}
involve  connections between fragmentation channels and  eigenchannels. 
 Eigenchannel method proved its versatility by extension to molecular spectra \cite{greene1985}.
 This formalism displays explicitly quantum numbers and parameters of atomic dynamics which makes it adequate for theoretical analysis of data.

{Zero of Jost function $J^-$associated with a bound state or resonance is a pole of collision matrix, $U \sim J^+ / J^-$, {\it eg} \cite{fano1986}.
The corresponding of Jost function in R-matrix terms $U \sim (1-L^* R)/(1-LR)$ is just the Siegert term $J^- \sim 1-LR$.
}
A  Jost matrix approach to multichannel electron scattering on coulombian fields  
is presented
in \cite{drukarev1978}, ch. 3.5, \cite{drukarev1978e}:
the Collision matrix for this problem is $I^*I^{-1} = (M - \lambda^*)(M - \lambda)^{-1}$. This Jost matrix $I$ is acting in space of fragmentation
 channels.
The matrices $M$ and $\lambda$ are logarithmic derivatives at channel radius of internal and channel wave functions. Observe they can be identified with $R^{-1}$ and $L$ of R- matrix theory. The corresponding Jost matrix for multichannel electron scattering is 
$I \sim R^{-1} - L$. 

All reaction channels, irrespective open  or closed, are treated  in same way both in above approach as well as in Eigenchannel method. Our MQD approach is based on effective Collision matrix which, at its turn, is related to Siegert equation displaying effect of closed channel on open ones. 

The comparison between Siegert state to well established MQD approaches is not an easy matter and it is not in streamline of this work; some of  above remarks do touch only   possible relationships.
One should note that Siegert state approach to MQD, in the present compact form,
has no direct application in analysis of experimental data. It is, mainly,  a methodological result and a physical demonstration  of role of Siegert state concept in MQDT.

\end{appendices}

\subsection*{Acknowledgements}
One of the authors (CH) acknowledges support of A v Humboldt Foundation and 
hospitality of the Munich University.








\end{document}